\documentclass[%
reprint,
superscriptaddress,
groupedaddress,
amsmath,amssymb,
floatfix
]{revtex4-2}

\usepackage{float}
\usepackage{fancyhdr}
\newcommand{\doil}{\textit{Digital Discovery}, 2024,  \textbf{3, }1221-1235}
\newcommand{\doir}{\href{https://doi.org/10.1039/D4DD00016A}{https://doi.org/10.1039/D4DD00016A}}
\fancypagestyle{firstpage}{
  \rhead{\doir}
}

\usepackage[normalem]{ulem}
\newcommand{\papertitle}{
Flexible, Model-Agnostic Method for Materials Data Extraction from Text Using General Purpose Language Models
}

\usepackage{graphicx}
\usepackage{dcolumn}
\usepackage{xcolor}
\usepackage{colortbl}
\usepackage{multirow}
\usepackage{makecell}
\usepackage{bm} 
\usepackage{hyperref} 
\usepackage[mathlines]{lineno}
\usepackage{float}
\usepackage{soul}
\usepackage{lipsum}

\begin{document}

\title{\papertitle}

\author{Maciej P. Polak}
\email{mppolak@wisc.edu}
\author{Shrey Modi}
\author{Anna Latosinska}
\author{Jinming Zhang}
\author{Ching-Wen Wang}
\author{Shaonan Wang}
\author{Ayan Deep Hazra}
\author{Dane Morgan}
\email{ddmorgan@wisc.edu}
\affiliation{Department of Materials Science and Engineering, University of Wisconsin-Madison, Madison, Wisconsin 53706-1595, USA}

\begin{abstract}

Accurate and comprehensive material databases extracted from research papers are crucial for materials science and engineering, but their development requires significant human effort. With large language models (LLMs) transforming the way humans interact with text, LLMs provide an opportunity to revolutionize data extraction. In this study, we demonstrate a simple and efficient method for extracting materials data from full-text research papers leveraging the capabilities of LLMs combined with human supervision. This approach is particularly suitable for mid-sized databases and requires minimal to no coding or prior knowledge about the extracted property. It offers high recall and nearly perfect precision in the resulting database. The method is easily adaptable to new and superior language models, ensuring continued utility. We show this by evaluating and comparing its performance on GPT-3 and GPT-3.5/4 (which underlie \texttt{ChatGPT}), as well as free alternatives such as BART and DeBERTaV3. We provide a detailed analysis of the method's performance in extracting sentences containing bulk modulus data, achieving up to 90\% precision at 96\% recall, depending on the amount of human effort involved. We further demonstrate the method's broader effectiveness by developing a database of critical cooling rates for metallic glasses over twice the size of previous human curated databases.

\end{abstract}
\maketitle
\thispagestyle{firstpage}
\section{Introduction}
\label{sec:introduction}

Obtaining reliable and comprehensive materials data is crucial for many research and industrial applications. If necessary information is not accessible through curated databases researchers typically must manually extract the data from research papers, a process that can be time-consuming and labor-intensive. Natural language processing (NLP) with general Language Models (LMs), and in particular, powerful large LMs (LLMs) trained on massive bodies of data, offer a new and potentially transformative technology for increasing the efficiency of extracting data from papers. These methods are particularly valuable when the data is embedded in the text of the documents, rather than being presented in a structured or tabular format, making it harder to find and extract. While currently LLMs often fall short in practical applications, struggling with comprehending and reasoning over complex, interconnected knowledge domains, they offer significant potential for innovation in materials science and are likely to play a crucial role in materials data extraction \cite{rev2}.

The rapid pace of development in NLP and frequent release of improved LLMs suggests they can be best utilized by methods which are easily adapted to new LLMs. In this paper we present such a flexible method for materials data extraction and demonstrate that it can achieve excellent precision and recall.

So far, the majority of materials data extraction approaches focus on fully automatic data extraction \cite{nlp_review,nlp_challenge,ner_review2,chemnlp,tshitoyan,rev_isayev}. Automation is clearly desirable, particularly when extracting very large databases. However, more automation tends to require more complexity in the software, sophistication in training schemes, and knowledge about the extracted property. In addition, if a high level of completeness is required from a database, the recall of these approaches may not be sufficient. In such fully automated approaches a large amount of focus has been placed on the complex task of named entity recognition (NER) \cite{ner1,ner2,ner3,ner4,matscibert}, so that the property, material, values and units can be extracted accurately. However, automatic identification of an improper recognition is still very challenging, which can reduce the precision of such approaches. Tools for automatic materials and chemistry data extraction, like OSCAR4 \cite{oscar4} or ChemDataExtractor \cite{cde,cde2}, have been developed and used to successfully extract large databases. A recent example includes a database of over 22 thousand entries for relatively complex thermoelectric properties \cite{cde_thermo}, at an average precision of 82.5\% and a recall of 39.23\%, or over 100 thousand band gap values \cite{cde_gaps}, with an average precision of 84\% and a recall of 64\%. More complex information such as synthesis recipes \cite{synthesis1,synthesis,synthesis2,synthesis2017,synthesis2019,synthesis2020} have also been extracted with automated NLP-based methods. Although not complete due to the relatively low recall, databases of that size are useful for training machine learning models \cite{ml2,ml1,cde_magnetic_ml,nlp_ml_synthesis,cde_ml,synthesis_ml,ml_mit}, and would be very time consuming or impossible to extract with virtually any other method than full automation. Other recent examples of databases created in a similar way include photovoltaic properties and device material data for dye-sensitized solar cells \cite{cde_solar}, yield strength and grain size \cite{cde_yield}, and refractive index \cite{cde_refractive,cde_refr_diel}. Other notable databases gathered with NLP-based approaches include more complex information than just data values, such as synthesis procedures \cite{synthesis,nlp_ml_synthesis}.
Recently, another method for structured information extraction, making use of the \texttt{GPT-3} capabilities was presented \cite{structured_data}. In that work, the focus is placed on the complicated NER tasks and relation extraction, at which \texttt{GPT-3} excels. In that approach, more complex sentences can be successfully parsed into structured information. A ''human in the loop'' approach was used to fine-tune the model, a technique that seems to be emerging as a method of choice to obtain higher performing models. Impressive performance was achieved in this work for structured information extraction, although at a price of a relatively large set of relatively complex training examples.

In addition, the emergence of highly specialized LLMs underscores the rapid advancement in the field. In Ref. \cite{honeybee} an instruction-based process specifically designed for materials science enhanced the accuracy and relevance of data extraction. Such specialized fine-tuning shows significant advantages in dealing with niche materials science tasks.

Recently, fully automated agent-based LLM approaches to analyze scientific text have been proposed as well, which are capable of answering science questions with information from research papers \cite{rev_agent1}, and generating customizable datasets \cite{rev_agent2}. Other fully automated LLM-based methods, including those that leverage complex prompt engineering workflows within LLMs have been proposed to curate large materials datasets of a higher quality than conventional automated NLP methods when used with state-of-the-art LLMs \cite{Polak2024}.

\begin{figure}
\centering
\includegraphics[width=.36\textwidth]{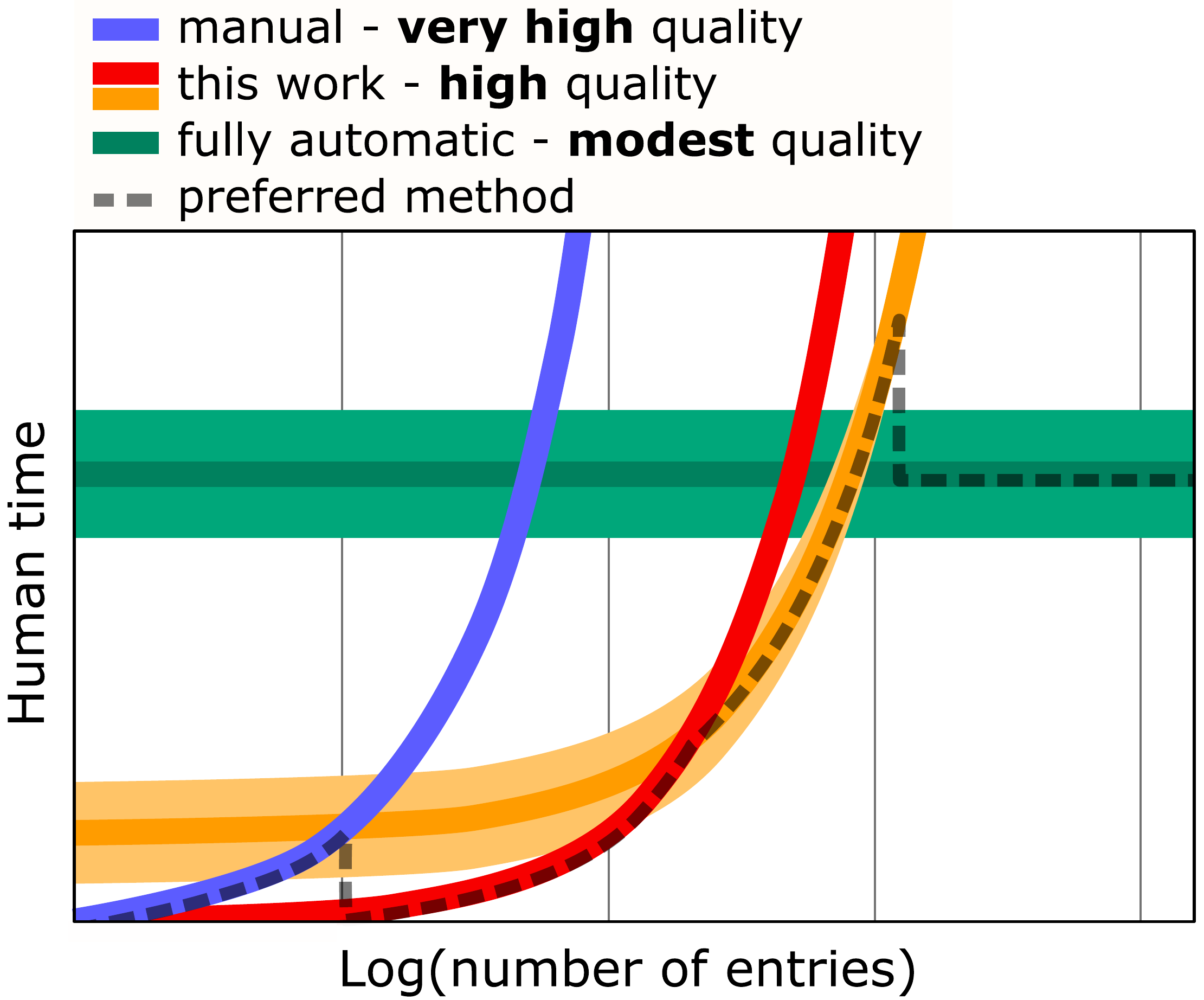}
\caption{Qualitative behavior of different types of approaches to data extraction, presented as human time required as a function of the size of the dataset. The broad range of the green (fully automatic), and orange (this work) represents the potential variation in the initial fixed time requirement, which may slightly influence the quality of the result. The dashed line suggests which method is the best choice for a given size of dataset.}
\label{fig:abstract}
\end{figure}

Depending on the nature of the data and the end goal for which the database is needed, there are different requirements for the resultant database and different optimal approaches for the data extraction. It is useful to organize methods along two broad axes. One axis is human time, which generally has the form 
\begin{equation}
\label{equation}
t=A + B\cdot N.
\end{equation}
Here $A$ is up-front fixed time to develop the method for a specific case and $B$ is a marginal time cost whose contribution scales with some measure of data quantity $N$ ($N$ represents some function of the number of papers, sentences, and data to extract). The other axis is database quality, generally represented by some combination of precision (what fraction of the found data are right) and recall (what fraction of the available data was found) of the database. A schematic plot representing required human time as a function of the size of the database for different methods, and the quality of their results, is presented in Fig.~\ref{fig:abstract}, where a logarithmic scale was used to emphasize the behavior for small-moderate sized databases, where the method is the most advantageous. While the classification of a dataset's size depends significantly on the context and scientific field, in this paper we base this classification, as detailed in the following paragraph, on an evaluation of the typical quantities of relevant materials data found in existing literature.

One limiting case, which we will call ''small data'', is one where only a small amount of data, up to around 100 points, is available in the literature (for example, properties that are very new, very hard to measure, or studied by only a small community), and where completeness and accuracy are highly valued. For example, as of this writing ''small data'' might refer to superconductors with $T_c > 200K$ \cite{supercond}.  It is typical to gather data for properties in the very small data limit fully manually, usually by experts in the field. Full manual curation is practical due to the limited number of papers and data and assures that the data is comprehensive and accurate. This fully manual approach is represented in Fig.~\ref{fig:abstract} as the blue line, which is preferred (dashed) for very small amounts of available data. Even though it is technically slower than other methods, even in the very small data range, it is still the method of choice due to the highest possible quality of the results.

The opposite limiting case, which we will call ''large data'', is when there is a lot of data in the literature, more than a couple thousand datapoints, the database is expected to be large, and modest precision and recall are acceptable. For example, such a database might be pursued for use for building machine learning regression and classification models on widely studied properties. For this large data case fully automated NLP-based approaches may be the most appropriate solution (see green curve in Fig.~\ref{fig:abstract} for large number of entries). However, such an automated approach can result in an incomplete database that may not be sufficient for certain research or industrial applications, e.g., where extremes of performance of just a few materials might be the primary interest. In addition, conventional, fully automated NLP approaches often require extensive retraining and building of parsers specific for different properties, as well as a significant amount of coding. These methods thus often require a significant initial investment of human time. 

Datasets in the middle between ''small'' and ''large'' are considered in this paper as mid-size, i.e. containing between around a hundred and a couple thousand datapoints.

The logic of the best approaches for these extremes is simple. Large data (e.g., $>~10^4$ data points) can be most efficiently extracted by spending human time on automating the extraction (leading to large $A$ and small $B$ in Eq.~\ref{equation}), and reduced precision and recall is often of limited consequence since so much data is available. Small data sets (e.g, $<~10$ data points) can be most efficiently extracted by spending human time on directly extracting the data (leading to small $A$ and large $B$ in equation Eq.~\ref{equation}), and high precision and recall is typically more important for smaller databases. However, the optimal approach for the middle ground between these scales, which represents many databases in materials, is not obvious.

We propose the use of a method that is most suited to creating these mid-size databases. With the recent significant advances in performance and availability of LLMs, there is opportunity for significant improvements by employing them as a part of a language processing workflow for the purpose of materials data extraction. This method uses a combination of LLMs methods, with some degree of human supervision and input, which allows one to relatively quickly extract data of high quality while at the same time requiring minimal coding experience and upfront fixed human time cost. The method leads to modest $A$ and $B$ in Eq.~\ref{equation}, making it better than human extraction or full automation in the medium-data scale range. Two variants of the method are represented in Fig.~\ref{fig:abstract} by the red and orange curves. They provide data of relatively high quality, approaching that of a fully manually created database, and scale well for medium sized databases. The proposed methods allow a database of up to the order of 1000 data points to be gathered in a few hours.

The general idea of breaking up the papers into sentences and classifying those sentences as relevant or not, perhaps with a model fine-tuned with human supervision is a commonly utilized language processing practice, including in materials science \cite{rev_1} . This general idea is also the core of the method presented here. However, we use a LLM to classify each sentence as relevant or not, parse each relevant category sentence with a LLM into a structured set of target data, and then perform human review of the extracted structured data for validation and fixing. The LLM classification is done either fully automatically (in a zero-shot fashion) or with some small human effort to fine-tune the LLM with example sentences. The LLM classification step typically removes about 99\% of the irrelevant data and leaves only about 1\% to be further analyzed, dramatically reducing human labor. The final human review is very efficient as only highly structured data is presented, and most are already correct or nearly correct. This method results in an almost perfect precision and recall for the resultant database, comparable to a fully human curated database, but at 100 times or less human effort.

There are three major advantages of this method compared to other possible approaches of data extraction with NLP. First, the method is very easy to apply, requiring almost no coding, NLP or LLM expertise and very limited time from the user. For example, the case where the LLM is provided by transformers zero-shot classification pipeline \cite{transformers} requires just 3 lines of code that are provided on the huggingface website. As another example, in the case where the LLM \texttt{GPT-3/3.5/4} is used, the API request is also just a few lines and provided to the user explicitly by the developers. Second, the method interfaces with the LLM through a standard classification task available in any modern LLM, making it possible to easily use the method with many present and likely any future LLMs. Thus the method can easily take advantage of the rapid improvements occuring in LLMs. Third, the method requires almost no knowledge about the property for which the data is to be extracted, with just the property name required for the basic application of the method.

In this paper we demonstrate the method by developing databases with multiple LLMs. The simplicity and flexibility of the method is illustrated by repeating the development of a benchmark bulk modulus sentence classification database with multiple OpenAI \texttt{GPT} models, including the recently released \texttt{GPT-3.5 davinci}, \texttt{GPT-3.5 (turbo)} and \texttt{GPT-4} \cite{gpt3,instructgpt}, as well as the \texttt{bart}- and \texttt{DeBERTaV3}-based language models \cite{bart, bart_benchmark,deberta,dbrt_xnli} hosted on huggingface, currently the most downloaded models for text classification. It is important to demonstrate the applicability and efficiency of the method on both simple, free, and accessible LLMs that can be easily used on a personal computer, and on LLMs which require significantly more computation and may be out of reach of some most people's resources for now. Even though there exist fully free and open LLMs, such as \texttt{OPT}\cite{opt}, \texttt{BLOOM}\cite{bloom} or \texttt{LLaMA}\cite{llama}, their use is computationally expensive and not convenient, which contradicts the spirit of ease and accessibility of the presented method. Therefore, we opted for The OpenAI's models whose API allows one to efficiently use the LLM on outside servers, although is not free. \texttt{GPT-3/3.5/4} are also currently the most popular LLMs, so its a choice that will likely be relevant for many users.

We demonstrate and benchmark the method on raw texts of actual research papers, simulating how the method will likely be used by science and engineering communities. We first assess the precision and recall of the method on a small set of papers and the property bulk modulus in order to demonstrate the excellent accuracy of the classification that can be obtained with this method. We then use the method to extract a modest sized but high quality database of critical cooling rates for metallic glasses.

The paper is organized as follows: Section \ref{sec:method} describes the approach in detail; Section \ref{sec:results} shows the results of benchmarks and statistical analysis of the obtained classification results for a bulk modulus sentences database; Section \ref{sec:discussion} discusses the developed database of critical cooling rates for metallic glasses, the possible utility of the method for purposes other than simple data extraction, and future possibilities in light of the rapid evolution of NLP methods and new LLMs. Section \ref{sec:datasets} describes in detail the benchmark bulk modulus sentence classification database used for assessment as well as the critical cooling rate of metallic glasses database.

\section{Description of the approach}
\label{sec:method}

\begin{figure*}
\centering
\includegraphics[width=.99\textwidth]{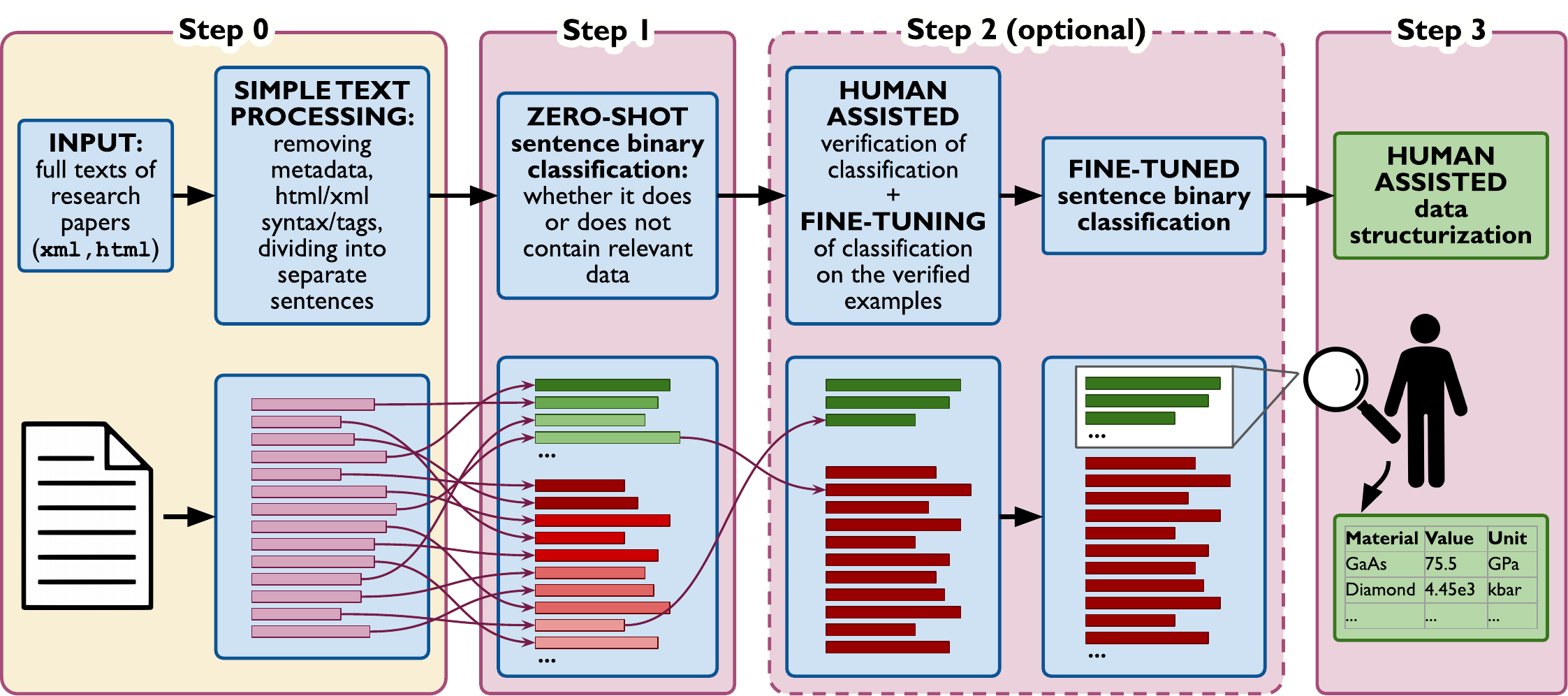}
\caption{A diagram of the steps necessary for NLP/LLM data extraction in the proposed method. The process starts with gathering and preparing the documents to be analyzed, a process not involving any NLP (Step 0), then a LLM is used to classify sentences by whether a sentence does or does not contain data for a given property (value and units) in a zero-shot fashion (Step 1). The pre-classified sentences are then (optionally) validated and used for fine-tuning the LLM and reclassifying the sentences with higher quality (Step 2). Finally the data is structured by a LLM/human assisted process, where the name of the material/system, the numerical value of the given property, its unit, and in some cases an additional detail, such as the temperature at which the value are obtained (Step 3). A detailed description of all steps can be found in Sec. \ref{sec:method}}.
\label{fig:schematic}
\end{figure*}

This section describes the steps involved in the method for data extraction from research papers. These steps are schematically presented in Fig.~\ref{fig:schematic}. Each step is first summarized (in bold), and then expanded to include details and observations we had during the work. Note that the first step is focused on gathering and basic processing of the papers and does not use any LLMs. Furthermore, this step (or one very similar) is present in any data extraction method and is not specific to our method. Since this step is largely a universal preprocessing that does not have any direct connection to the method we label it the zeroth step, thereby allowing the first step to denote a step directly connected to data extraction.

\subsection*{STEP 0:}
\label{sec:step0}
\noindent\textbf{Obtaining and postprocessing the raw html/xml paper texts into human-readable format.}

\paragraph*{\textsc{Input:}} This step starts with gathering the papers for analysis (e.g. from ScienceDirect API \cite{sciencedirect}) in an \texttt{xml/html} format. This usually involves searching for papers through a query to the publisher's search engine, and simply downloading every paper that comes up as a result. If at this stage any relevant papers are missed, the data will not be extracted, so it is safer to use a broad search query or combine results of multiple queries. Including additional data does not increase the amount of human time, only involves more processing for the NLP model. This step takes very little human time, does not depend on the size of the database and is already mostly automated through the publisher, since only the query is required to obtain a list of matching documents.

\paragraph*{\textsc{Simple text processing:}}
After the papers are downloaded the metadata and \texttt{html/xml} syntax is removed. We keep all the paragraphs and the title and remove the rest of the content. Then, we remove the \texttt{html/xml} markup syntax and tags. At this point all that is left is pure text. This cleaned up text is then split into separate sentences, according to regular rules for how sentences are terminated. At the end of this step we are left with the raw data that may be fed to the LLM and analyzed. There is no need for any human evaluation of the data at this point. Whether an entire paper is unrelated, or some of the paragraphs contents are, it will simply be analyzed by the LLM and deemed irrelevant. This step takes very little human time, and the amount required does not depend on the size of the database or the extracted property. The exact method for removal of \texttt{html/xml} syntax and splitting into sentences can vary. It can be done by text processing through regular expressions (an example can be found in the codes, Sec.~\ref{sec:data_availability}), or ready-made specialized python libraries and their functions (such as \texttt{lxml}, \texttt{nltk.tokenize.sent\_tokenize} \cite{NLTK,lxml}), depending on the user preference. 

It is important to note here that further simple text processing of the cleaned up text to keep only plausible sentences, e.g. using regular expressions to keep sentences with easily identified essential information, can, and probably should be performed at this point. Although such an additional processing step does not influence the method or the final outcome and quality of the produced database, such processing can significantly reduce the amount of data to be categorized by the LLMs. This simple processing will certainly reduce the compute time needed for the LLM and can reduce costs if the LLM is not free. How this text processing is performed depends on the task and the amount of knowledge about the data or property to be extracted. For example, if we know the data is numeric we can keep just sentences containing a number. In the case of bulk modulus  (see Dataset 1 in Sec.~\ref{sec:datasets})), keeping only sentences containing a number cuts the amount of data to be processed in half and does not lower recall (i.e., keeps all relevant sentences). If some amount of knowledge about the quantity to be extracted is available it can be used to further select the most promising sentences. For example, if we know the expected units of the data we can further process the remaining sentences to keep only those that contain such units. In the case of bulk modulus (see Dataset 1 in Sec.~\ref{sec:datasets})) keeping only sentences containing units of pressure (pascals and bars with possible metric prefixes, $N/m^2$), lowers the amount of possible candidates to less than 20\% of the initial set, still without any loss of recall. Such refinements can be continued to further narrow down the search, but each subsequent step relies on a deeper knowledge of the property in question and increases the risk of reducing recall. In the work presented here we assume the most demanding situation for the method, in which no prior knowledge of the property is assumed. Therefore we only narrow down the search to sentences containing numerical values.

\subsection*{STEP 1:}
\label{sec:step1}
\textbf{Zero-Shot binary classification of sentences to produce unstructured data, i.e. set of sentences containing values for a given property. The classification puts sentences in two categories: \emph{positive}, which are sentences containing the data (numerical value for a given property with its corresponding unit), and \emph{negative}, which are sentences \emph{not} containing the data.}\\
Depending on the LLM used, the zero-shot may require as little as just the name of the desired property as the label of the class (name of the property), or require a full prompt phrase (e.g., \texttt{GPT-3/3.5/4}). Since the most recent and powerful LLMs make use of a prompt (e.g., \texttt{GPT-3/3.5/4}), we focus on that case here. The prompt (a single set of words, typically a phrase that makes grammatical sense) given to the model plays an important role. The impact of prompts has already been widely observed in NLP-based text to image generation tools (e.g., DALL-E2 \cite{dalle}, MidJourney \cite{midjourney}, Stable Diffusion \cite{stablediffusion}) and a similar situation occurs in the present application. Depending on the completeness and phrasing of the prompt, the results for classification may be dramatically different. In our experience, however, prompts that do not contain false and misleading information almost always result high recall, and it is mainly the precision that is affected. In addition, more complex prompts do not necessarily guarantee a better result and may not be necessary.
It is worth mentioning that with modern LLMs, other approaches such as one/few-shot (providing a prompt together with one or a few example outputs) or even more complex ways of extracting data, involving multiple subsequent prompts, have been shown to be very effective \cite{gpt3}. In this work, however, not only is the zero-shot efficient enough for the classification task, but it is also the simplest, most straightforward to apply and assess its performance, and ensures higher flexibility and transferability to other properties, so the other more complex methods have not been explored.

Fig.~\ref{fig:zeroshot-comparison} shows the zero-shot result statistics for the different models, including \texttt{GPT-3.5} (whose technical names are \texttt{text-davinci-002} and \texttt{text-davinci-003}) and other \texttt{GPT} models including 3.5 (turbo) and 4, which underlie \texttt{ChatGPT}. The \texttt{Chat} models do not output probabilities so full precision recall curves cannot be plotted, only a single point, which for all \texttt{Chat} models has 100\% recall. The \emph{p1} and \emph{p2} stand for two different prompts.\\
\emph{p1: Does the following sentence contain the value of bulk modulus?}.\\
\emph{p2: A sentence containing bulk modulus data must have its numerical value and the units of pressure. Does the following sentence contain bulk modulus data?}\\
Only the first token of the model's response was evaluated and in all cases it was either a ''yes'' or a ''no'' (case-insensitive), as expected, allowing for an unambiguous classification. As an example consider the two following sentences:\\
\emph{1. After full lithiation, the phase transformed to Li13Sn5, which has the bulk modulus of 33.32 GPa and the Poisson's ratio of 0.205.}\\
\emph{2. The structure of polycrystalline copper is cubic with lattice parameters a = b = c = 3.6128 (1) Å at 0.0 GPa.}\\
We would get a "yes" response for the first, and a "no" response for the second.\\
Even though \emph{p2} contains more seemingly valuable information, it did not necessarily perform better. We experimented with various different prompts, and straightforward prompts similar to \emph{p1} performed most consistently and predictably for most models. Therefore a simple prompt: \emph{Does the following sentence contain the value of [name of property]?} is our strong recommendation. The one exception is \texttt{GPT-4}, where a more detailed prompt resulted in a significantly better result. This is due to an improved accuracy of prompt interpretation and following the prompt instructions in \texttt{GPT-4}.

It is worth noting that some models, such as \texttt{GPT-3/3.5 davinci} and \texttt{GPT-3.5/4} (chat) at the time of writing this article are not free to use. Therefore, the flexibility to use different LLMs within the method is very valuable, as some free models, while not necessarily capable of accurately performing the more complex tasks such as automated data structurization, and although overall generally less capable than \texttt{GPT}-based models, perform well enough in the simple task of classification to produce satisfying results. However, in the case of OpenAI \texttt{GPT-3}, both model usage and fine-tuning is done on outside servers, so in a situation where computational resources are not available to run locally, it may enable one to use the best models at a low cost.

\subsection*{STEP 2 (optional):}
\label{sec:step2}
\noindent\textbf{Human assisted verification of the zero-shot classification of sentences. The highest scoring unstructured data (most likely to be a true positive), pre-classified in Step 1, is manually classified into \emph{positive} and \emph{negative} sentences to provide a new training set of sentence for fine-tuning the classification process. The new training set is then used to fine-tune the model and classify the sentences again to obtain higher precision and recall.}\\
This optional step is just a chance for the human user to provide confirmation or correction to particularly important zero-shot classification data from Step 1 and then use those checks to fine-tune the LLM. Similar steps are often taken in other data extraction approaches, and machine learning in general \cite{structured_data, rev_1, ml_learn}. Specifically, as the highest scoring sentences are being manually verified, a new training sets consisting of true positive and true negative examples is built. Since the precision of results of Step 1 is typically around 50\% at 90\% recall (see Fig.~\ref{fig:zeroshot-comparison}~(d)), the created sets are typically close to equal in size. The human labeled sets consist of \emph{positive} cases, which represent \emph{true positives} from Step 1, and \emph{negative} cases, consisting of \emph{false positives} from Step 1. The latter are the most valuable counter-examples for the negative training set, as these are the sentences the easiest for the model to confuse for positives. If after reaching the desired amount of verified positive sentences the corresponding set of negative sentences is smaller, it may be complemented with random sentences from the analyzed papers (the exceeding majority of which are negative). Fig.~\ref{fig:learning} shows how the classification model improves when fine-tuned on datasets of increasing size. A detailed analysis of that figure is present in Sec.~\ref{sec:results}, where a conclusion is made that after around 100 positive sentences for the quicker learning models such as \texttt{GPT-3 davinci/GPT-3.5} or \texttt{bart}, we start to observe diminishing returns with this human labeled dataset size increase and it may not be worth spending more human time on obtaining more examples. Therefore, we recommend to perform the manual verification of the zero-shot classification until 100 positive sentences (and a corresponding 100 negative - made easy due to the ~50\% precision) are obtained, a number easy to remember and satisfactory for an efficient fine-tuning dataset.

This steps usually takes no more than 30 minutes for approximately 100 sentences - each sentence has to be classified only in a binary fashion which is a very simple task and takes just a few seconds per sentence. The classification is as simple and straightforward as assigning $1$ for positive and $0$ to each sentence in a spreadsheet. The fine-tuning itself, for the small locally hosted models (\texttt{bart} and \texttt{DeBERTaV3}), takes around 30 minutes on an older workstation CPU (Intel(R) Xeon(R) CPU E5-2670), 20 minutes on a modern laptop CPU (Intel(R) Core(TM) i9-9880H), and can be reduced to just a few minutes if GPUs are used. The OpenAI models are fine-tuned on external OpenAI servers in less than 30 minutes and do not require any local resources.
After this step is performed and the sentences are once again reclassified using the now fine-tuned model the precision and recall are greatly improved, as can be seen in Fig.~\ref{fig:learning}.

This Step 2 is optional and is generally done to improve the quality data collected from Step 1. Improving precision of data at this stage will reduce the human time needed in data structurization in Step 3 (see Sec.~\ref{sec:step3}) to review the data. However, for small datasets the human time in Step 3 is very modest, and this Step 2 may not be worth the extra effort. Thus whether it is performed or not typically depends on the size of the dataset. For small datasets, and if a recall of around 90\% is satisfactory this step can be entirely omitted. As seen in Fig.~\ref{fig:zeroshot-comparison}~d), the precision at 90\% recall after Step 1 is over 50\% for the best models, which means that for every true positive sentence, there is only one false positive - a reasonable number to be removed by hand during data structurization (Step 3). For small datasets, up to a few hundred values, verifying around 100 positive sentences to perform additional fine-tuning to improve the precision may turn out to be more labor intensive that proceeding straight to data structurization, and improving the precision manually by simply ignoring false positives.
It is crucial to understand that the recall obtained at this step (or that has been obtained after Step 1, if this optional step is skipped) will be the recall of the final database, while the precision will be improved to near perfect in the next step (Step 3).

\subsection*{STEP 3:}
\label{sec:step3}
\noindent\textbf{Data structurization (template filling). In this step, extraction of the structured data is performed. Here, by structured data we mean the full information necessary to provide a datapoint: the name of the material/system, the numerical value of the given property, its unit, and in some cases an additional detail, such as the temperature at which the value was obtained. At the same time as the data is extracted the precision of the result is improved to perfect (or near perfect, depending on the expertise and accuracy of the human supervising this step), by simply ignoring the false positive sentences left over from previous steps. The result of this step is a final, curated structured database.}

The user will typically perform this step by first ranking the sentences by their probability of being relevant (classification scores in the case of small LMs, \texttt{bart} and \texttt{DeBERTaV3}, or log probabilites in the case of GPT-3), which is the output from Step 1 (or Step 2 if performed), and start reviewing the list at the top, working down until they decide to stop. As the user works through the results in that fashion, they traverse down the precision recall curve (PRC) (see. Figs.~\ref{fig:zeroshot-comparison}~(a) and \ref{fig:learning}~(a)). While the recall is impossible to assess without knowing the ground truth, the user is fully aware of the precision of the data they have already analyzed, therefore using the PRC they can estimate the recall and decide to stop when a desired recall is reached (with the assumption that the PRC are similar to those shown in Figs.~\ref{fig:zeroshot-comparison}~(a) and \ref{fig:learning}~(a)). For best models, reaching recall of around 90\% (close to that of a fully manual data curation) without performing the optional Step 2 happens for a precision close to 60\%, while for a fine-tuned model (with Step 2), for a precision over 80\%. It is entirely up to the user to decide the quality they require from their database, and the quality of the results will be proportional to the amount of time spent in this step. Recall of 90\% seems to be a reasonable value to stop the process, as the precision sharply drops for higher values, which diminishes returns for the human time involved. However, this behavior may vary depending on the case, which will be discussed further in the Sec.~\ref{sec:results}. 

In general, human assisted data structurization, even when only the sentences containing the relevant data are given, may be a tedious and time consuming task. However, at this point it is the only method that can guarantee an almost perfect precision. For an inexperienced user, extracting one datapoint from a given sentence and its surrounding context fully by hand may take as long as 30 seconds, depending on the complexity of the property being analyzed and how it is typically expressed in research papers. Considering this, only relatively modest sized databases are reasonable to create. However, with experience, this time quickly reduces as the user gets used to the process. In addition, more experienced users may employ simple computer codes, e.g. based on regular expressions, which would preselect possible candidates for values and units, reducing the time significantly. In the longer term, it is likely the NLP tools will help automate this data structurization step. Some models, like GPT4, offer structured format output, such as json, which may be used to assist the final data extraction step. However, they do not do this very effectively at present without either human supervision or a major effort to tune them. For example, \texttt{GPT-3/3.5/4} is capable of parsing unstructured data in a zero-shot fashion, with no need for retraining. In the case of our bulk modulus sentences dataset we found that in over 60\% of cases \texttt{GPT-3/3.5/4} is capable of correctly providing the entire data entry for a given property (name of the material/system, value, unit), and an incomplete datapoint (wrong material/system, but correct value and unit) in over 95\% of cases. The only drawback that prohibits a full automation of this step with a LLM is the inability to automatically and unambiguously distinguish between correct and incorrect extracted datapoints. Even though the model does not tend to make up (hallucinate) data, it sometimes provides an incomplete or inaccurate extraction (e.g. "alloy" instead of "AlCu alloy" for the material, or "100" instead of "greater than $/>$ 100" for the value, etc.). However, human assistance in determining whether the data has been structurized properly, and in case it was not, fixing it by hand, can easily remedy that problem. Since almost all values and units are extracted correctly, and only less than half of the material names require fixing, using a LLM approach greatly reduces the human time and effort required for data structurization. Using an LLM we found that the average human time required to extract each good datapoint was reduced to under 10 seconds, keeping the same, almost perfect precision. Thus, and NLP-assisted data structurization, while still a tedious process, enables one create databases of up to around 1000 entries (more or less, depending on the users predisposition to and efficiency at repetitive tasks), in one workday. This timing includes the whole process, beginning (Step 0) to end (structurized database after Step 3), although almost all the human time is spent in Step 3.

While the value, units and the optional additional details most often occur within the positive sentence, the name of the material is often missing from that sentence (sentences are often similar to e.g. \emph{We determined the bulk modulus to be 123 GPa.}). In those cases the system is described most often in the preceding sentence, and if not, then in the title of the paper. In a vast majority of cases (96\% in our bulk modulus dataset) the full data information is available to be extracted from a sentence, that preceding it, and the title, so we do not search for it in other places. In the rare case when the full datapoint cannot be extracted, we record an incomplete datapoint. We also note that even in NLP models finely tuned for structurized data extractions, the further apart the relevant data are from each other, the more difficult it is for the model to accurately extract the relevant data, so those datapoints would very likely be incomplete with other NLP-based approaches as well.

\section{Results}
\label{sec:results}

\begin{figure*}
\centering
\includegraphics[width=.875\textwidth]{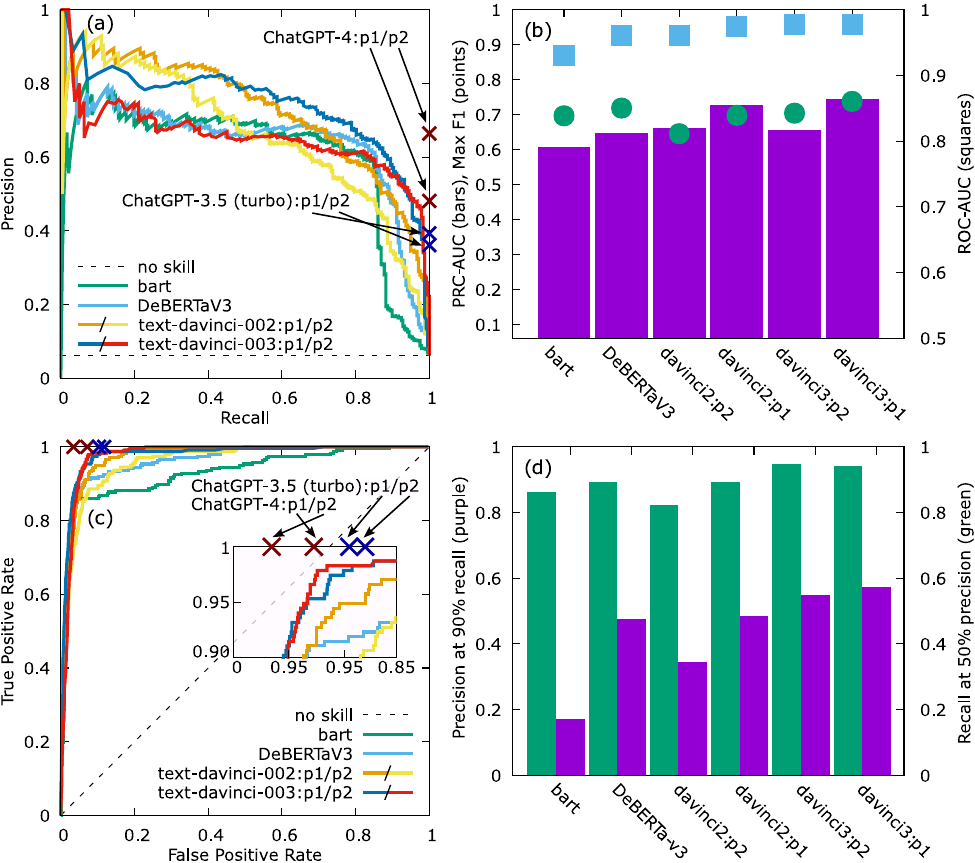}
\caption{Performance of different models after Step 1 (zero-shot binary classification of relevant sentences based on whether they contain bulk modulus data). (a) precision recall curves, (b) area under precision recall curve (PRC-AUC) (bars), maximum F1 score (circles), and area under receiver operating characteristic curves (ROC-AUC) (squares, right y-axis), (c) receiver operating characteristic curves with an inset  the upper left corner, (d) precision at 90\% recall and recall at 50\% precision (right y-axis). The \texttt{no skill} line represent a baseline model where the classification is random. \texttt{Chat} models do not output probabilities, therefore only one point of the curves in (a) and (c) is available for each \texttt{GPT-3.5} (chat) and \texttt{GPT-4} models and is labeled with dark blue and dark red  $\bm{\times}$ respectively. Note that all \texttt{Chat} models have 100\% recall. Labels in panels (b) and (d) have been shortened, but represent the same models as those in the legend of (a) and (c). \texttt{p1} and \texttt{p2} in the \texttt{davinci} models represent two different prompts (see Sec.~\ref{sec:step1}).}
\label{fig:zeroshot-comparison}
\end{figure*}

\begin{figure*}
\centering
\includegraphics[width=.8\textwidth]{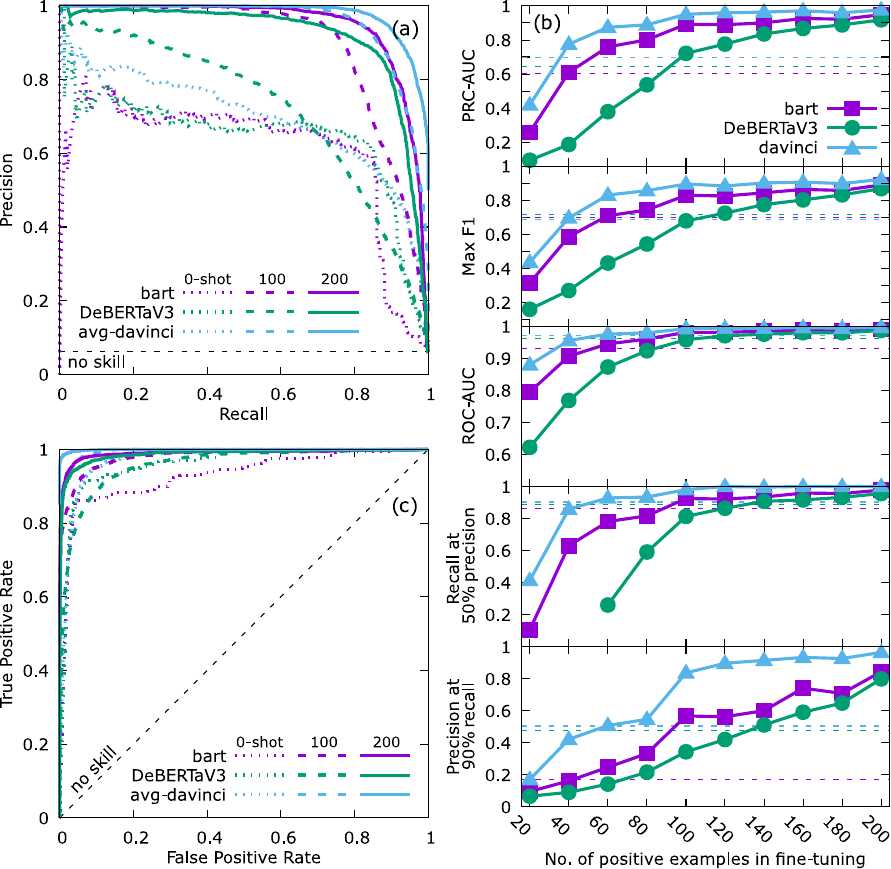}
\caption{Comparison of performance of different methods after fine-tuning (Step 2, Sec.~\ref{sec:step2}, binary classification of relevant sentences based on whether they contain bulk modulus data). Panel (a) shows precision recall curves, dotted lines correspond to the zero-shot (0-shot) result (\texttt{davinci} are averaged into one as described in the text), dashed and solid line correspond to fine-tuning on 100 and 200 positive sentence examples, respectively, (b) learning curves, i.e. performance metrics as a function of training set size, top to bottom: area under precision recall curve (PRC-AUC), maximum F1 score, area under receiver operating characteristic curves (ROC-AUC), recall at 50\% precision, and precision at 90\% recall. The horizontal thin dashed lines in corresponding colors represent zero-shot results. (c) receiver operating characteristic curves for the same data as in (a).}
\label{fig:learning}
\end{figure*} 

\textbf{This section provides a comprehensive analysis of various language models' performance in classifying relevant sentences. The analysis highlights the superior performance of the GPT family of models in a zero-shot approach and demonstrates the effectiveness of fine-tuning, while also discussing the results in the context of the accessibility of different models. It also addresses the challenges posed by highly imbalanced datasets and discusses strategies for reducing human effort in data processing.}

Fig.~\ref{fig:zeroshot-comparison} summarizes the result from Step 1 in Sec.~\ref{sec:method} for a bulk modulus analysis. The papers, sentences, ground truth category statistics, and other information is provided in Sec.~\ref{sec:datasets}.  The ground truth for Step 1 was determined by human labeling. The following precision recall curves (PRCs) and receiver operating characteristic curves (ROCs) are constructed in the usual way, which is by plotting the relevant metrics while varying the cutoff used for the lowest value of probability accepted as a positive classification for sentence relevance. Panel (a) shows a PRC for the models tested in this paper. The two different curves for each of the \texttt{GPT-3.5 davinci} and \texttt{GPT-3.5/4 (chat)} models correspond to two different prompts used in classification (see Sec.~\ref{sec:step1}). All of the tested models perform similarly, with \texttt{bart} struggling slightly more than others in achieving higher recall. The \texttt{ChatGPT} models result in only a single point, as the probability is not output from these models. All \texttt{Chat} models result in 100\% recall, with \texttt{GPT-3.5} (chat) performing similarly to base \texttt{GPT-3} models, which was expected since they are a part of the same family of models and based on similar architecture. The next generation \texttt{GPT-4} performs better, in particular with a more informative prompt (\emph{p2}, see Sec.~\ref{sec:method}). This is a result of an improved instruction-following capabilities of the \texttt{GPT-4} and a higher capability to apply knowledge provided in the prompt when producing results, which suggests that further prompt engineering may provide an even more improved performance in zero-shot classification in this, and likely in future, LLMs. However, this better performance of \texttt{GPT-4}, although impressive, ultimately may still be eclipsed by the even better performance of the fine-tuned \texttt{GPT-3 davinci} model (discussed later), and its significantly lower cost. A more quantitative measure of models' performance is presented in panel (b), where the area under the precision recall curve (AUC-PRC) alongside a maximum F1 score are presented. The \texttt{GPT-3.5} models, in particular using the first prompt (p1) show the highest scores, while \texttt{bart} and \texttt{DeBERTaV3} rank lowest in PRC-AUC. It is important to notice, however, that the datasets analyzed here are heavily imbalanced, with negative results outnumbering positives by more than 2 orders of magnitude. This places the naive \emph{no skill} in Fig.~\ref{fig:zeroshot-comparison}~(a) line, representing an entirely random model, close to zero (as opposed to at 0.5 for a fully balanced set), lowering the entire PRC compared to a balance set. Fig.~\ref{fig:zeroshot-comparison}~(c) shows the ROC, which is insensitive to dataset imbalance, and shows much higher AUCs (panel (b)) than those of PRCs. The conclusions from ROCs are similar to those from PRCs; \texttt{GPT-3.5/4} performs best, with \texttt{bart} scoring lowest, while still performing reasonably well. A non-LM approach based on regular expressions was also evaluated for comparison. In the case of bulk modulus sentences, a simple regular expression (regex) capturing sentences containing any number (\texttt{[0-9]}), the case-insensitive phrase "bulk modulus," and units of pressure (\texttt{[MG]*Pa|kbar}) resulted in an 82\% precision and 72\% recall (F1=0.76). While this result is comparable to the maximum F1 of zero-shot results of smaller LMs, LLMs such as GPT-4, as well as fine-tuned models, perform noticeably better. In addition, regex-based approaches do not directly offer a precision-recall curve, which would allow adjusting the balance to maximize recall without significantly sacrificing precision (see Sec. \ref{sec:step3}). Even though chat models such as GPT-4 do not offer the precision-recall curve either, in our test they performed at 100\% recall, so this fact was irrelevant.

It is informative to consider the implication of the ROCs and PRCs for the efficiency of the human effort in our method. The step that requires most of the human time for a modest size database or larger (e.g., a few hundred entries or more) is Step 3, where the user must read and structure output from each sentence categorized as positive in Step 1 (or Step 2 if used). In Sec.~\ref{sec:step3} we suggested that the user limit their review of sentences in Step 3 unless a desired recall (implied by precision through the PRC) is achieved.

In some applications one might wish to target a high recall irrelevant of the human time required in step 3. To give a sense of how that might impact the method, Fig.~\ref{fig:zeroshot-comparison}~(d) and \ref{fig:learning}~(b) show the precision for 90\% recall after Step 1 and after the optional Step 2, respectively. Consistent with our above discussions, the best models can achieve this recall with more than 50\% precision using even just the zero-shot approach (Step 1). For less robust models, a 50\% precision requires tuning (Step 2). For the worst models and using just zero-shot learning, the precision is about 17\%, meaning the user would be extracting useful data from only about 1 in every 6 sentences reviewed. This would likely still be practical, but could become very tedious for a database of even a few hundred final entries. However, the important implication is that if one uses the best models (\texttt{GPT-3.5/4}), even a quite high recall requirement, e.g., 90\%, can be achieved using very efficient sentence review, with almost every (more than 90\%) sentence presented to the user containing relevant data.

Fig~\ref{fig:learning} demonstrates how the performance of each of the models is improved if the optional fine-tuning in Step 2 is performed, as a function of the size of the training set. Panel (a) shows PRCs before fine-tuning (zero-shot) and compares them to PRCs after fine tuning on 100 and 200 positive sentences. While all models eventually show improvement, fine-tuning is clearly the most beneficial for the \texttt{GPT-3 davinci} (note that currently only the older generation \texttt{GPT-3 davinci} is available for fine-tuning). Similarly, various metrics describing the quality of the model are presented in Fig.~\ref{fig:learning}~(b), where learning curves as a function of the size of the fine-tuning training set are shown. The the x-axis represents the number of positive sentences included in the training set (with an assumed equal number of negative sentences). The shape of the learning curves differs for different models, with \texttt{GPT-3 davinci} model performing best (i.e. achieves higher performance metric values for smaller training sets) and learning the quickest (i.e. converges closer to best observed performance metric values for smaller training sets), \texttt{bart} following second, and \texttt{DeBERTaV3} third, across all metrics.
One may notice that performance of the fine-tuned models trained with very small training sets perform worse than zero-shot (Fig. \ref{fig:zeroshot-comparison}). When the model is fine tuned on a very specific and not very diverse set of information, the model's weights are updated with information inadequate to constrain it resulting in less accurate performance.
For \texttt{davinci}, slope starts to decrease rapidly (curve starts to saturate) for as few as 60-80 positive sentences in the training set, for \texttt{bart} that occurs at around 100 positive sentences, and for \texttt{DeBERTaV3} closer to 160. 
Even though not all of the curves are fully saturated for the above mentioned dataset sizes, constructing larger fine-tuning training sets is likely to waste more human time than it is going to gain in Step 3. Our recommendation, if the optional step 2 is performed, is to initially use a training set of around 100 positive sentences and the \texttt{GPT-3 davinci} model or the smaller and free \texttt{bart}. This size of 100 positive sentences is very manageable to obtain with human-assisted verification of classification after Step 1, and typically does not take more than 30 minutes. It is worth noting that although we expect this number to be transferable to other properties it has not been verified thoroughly on other properties.
Whether to perform the optional Step 2 (fine-tuning) will ultimately depend on the size of the database. As mentioned before, for larger databases, this improvement will be beneficial and save overall human time needed to curate the database by making Step 3 more efficient, while for small databases, up to a couple hundred datapoints, the time spent on the fine-tuning in Step 2 might be more than is saved during the data structurization in Step 3.

\section{Discussion}
\label{sec:discussion}

\textbf{The paragraph discusses the practical application of the presented approach to curate an extensive and accurate database of critical cooling rates for metallic glasses by analyzing a large volume of scientific literature. Comparison to existing, manually curated database and other automated methods is provided. Utility for complex data-oriented tasks like machine learning and the method's potential to handle unrestricted searches effectively is then discussed.}

To provide an example use-case for the method, we applied it to curate a high quality and highly accurate database of critical cooling rates for metallic glasses (Sec.~\ref{sec:dataset_critcool}). 668 papers responded to the query ''bulk metallic glass''+''critical cooling rate'', which is more than what a human researcher would be analyze manually in a reasonable timeframe. The proposed method resulted in 443 datapoints consisting of the value of materials, their critical cooling rate, and the unit in which they were expressed in the paper. These results, include all mentions of critical cooling rates, with different degrees of specificity, e.g., accurate values for specific compositions (the ideal result), value ranges for specific materials, and value ranges for broad families of materials. The obtained database covers the range of expected values very well, with values ranging from $10^{-3}\ K/s$ for known bulk metallic glass formers, to $10^{11}\ K/s$ for particularly bad glass formers. The well known Pd-based bulk metallic glasses ($Pd_{43}Cu_{2}Ni_{10}P_{20}$ and $Pd_{43.2}Ni_{8.8}Cu_{28}P_{20}$) are identified as those with the lowest values of critical cooling rates, while simpler alloys such as AgCu, PdNi or NiBe and pure metals such as Co are identified as those with the highest critical cooling rates, which further validates the results.
The obtained data, cleaned up for direct use in data oriented tasks (such as machine learning) i.e postprocessed to only include unique values for uniquely specified systems yielded 211 entries. Within these, 129 are unique systems (multiple values are reported for some systems and we kept these to allow the user to manage them as they wish). The database is larger than the size of a recently published manually curated database of critical cooling rates \cite{ben_bmgs}, which is the most state-of-the-art and complete such database of which we are aware, and consists of only 77 unique compound datapoints.
To provide comparison to other existing methods, we used ChemDataExtractor2 (CDE2) \cite{cde2}, a state-of-the-art named entity recognition (NER) based data extraction tool. With CDE we obtain a recall of 37\% and precision of 52\%, which are comparable to those reported for thermoelectric properties (31\% and 78\%, respectively) obtained in Ref. \cite{cde_thermo}.

Searching for a given property does not typically add any restrictions on the search other than the property itself, i.e., the search is unrestricted. In the case of the method proposed here, unrestricted search will identify and help extract all datapoints for the target property from the input set of documents. Therefore, if the user desires a database limited to, for example, a given family of systems, the limitation would have to be enforced in some additional way. This constraint could be done by limiting the input set of documents through a more strict search query, but even that does not guarantee that only the desired values will be extracted, as many papers mention a wide range of results, even if technically focused on a particular topic. Limiting the final database can be easily done manually in Step 3 (Sec.~\ref{sec:step3}), but depending on the property and the size of the desired subset, limiting the data at that stage may take a lot of human time and be inefficient. In principle, more restrictions than just the property can be imposed on the NLP level, but such abstract concepts as families of materials are very challenging even for the best LLMs and greatly reduce the quality of the zero-shot results (Sec.~\ref{sec:step1}) and would require significantly more training in (Sec.~\ref{sec:step2}).
This problem is highly dependent on the property in question. For example, an unrestricted search for critical cooling rates while limiting the search in Step 0 to papers responding to a query ''bulk metallic glasses''+''critical cooling rate'' was quite effective for our goals of obtaining all ranges of critical cooling rates for metallic glasses. But if one wanted, say, an overpotential for water splitting, restrictions on many factors, e.g., temperature or pH, might be essential to obtaining useful result and difficult to screen on in the initial Step 0.

A particular example of where unrestricted searches can be problematic occurs when searching for properties which are relevant in many fields when one is interested in only a particular field, and/or which have many possible associated restrictions which are needed to make the data useful. A specific example of this problem occurred for us when we explored constructing a database for ''area specific resistance'' (ASR) for anode materials of proton conducting cells. In step zero we searched for ''area specific resistance''+''proton conducting fuel cells (and similar terms)'' The method proved very successful at identifying sentences containing ASR and structuring the data, as it was asked to do. However, the method captured ASR in a wide variety of contexts, including single phase and composite materials, porous and non-porous materials, electrodes and electrolytes, steels, interconnects, coatings, varying temperatures, and ASR in both fuel cell and electrolysis operation modes. To obtain a simple and immediately useful dataset we were interested in single phase dense anodes operating in fuel cell mode with temperature information. Imposing such limitations was dramatically harder than the basic data extraction. Although one might have different goals than the ones just mentioned, it is very unlikely that one is interested in gathering information for all of the above data in a single database. Restricting the set of input documents was able to help to a certain degree to move the balance of the obtained results in the desired direction, but did not solve the issue entirely. From such a wide variety of contexts, identifying only those we were interested in required a relatively deep knowledge from the person performing the data extraction and required significantly more human time to extract than in case of datasets where the property is more uniquely identified. In fact, we stopped developing this database due to these many challenges, although for someone willing to commit 4-5 days of human time in step 3 the desired database is certainly practical to develop.

On the other hand, the lack of restrictions in the model may have other benefits, as it expands the possibilities of the kinds of information that can be extracted. For example, the method can be used to extract many kinds of information, not just property values. Step 1 with models like \texttt{GPT-3 davinci/GPT-3.5} or \texttt{GPT-3.5/4} (chat) broadly describe the type of text we are looking for, and Step 2 fine-tunes to better classify the relevant sentences. While we utilized this classification search to find sentences containing numerical data for a given property within text paragraphs of research papers, data may be present in other places such as tables or figures. The classification approach can be easily used to search for non-textual data such as tables or figures containing the relevant information, by classifying their captions. In case of a positive table classification, it would be followed by manual or algorithmical extraction from the already structurized table. Furthermore, classification can be used for more abstract concepts, such as suitability of a given material for a certain application, personal opinions of authors about promising directions of future research, or any other concept that can be characterized as a group of example texts for the model to train on, and classify in a binary fashion.

It is also important to remember that the method we present here is not restricted to the LLMs explored in this paper, and is in fact designed to be quickly adapted to new and improved LMs.

\section{Summary}
We have shown a simple and efficient method for materials data extraction from research papers. The simple concept of binary text classification using a LLM is involved as a key step in the method, which allows for a high flexibility in the language model used as virtually all modern language models are very capable at text classification. We determined \texttt{GPT-3/3.5/4 models} to be the best performers, but evaluated other, less expensive and more accessible alternatives such as \texttt{bart} or \texttt{DeBERTaV3}. By including a highly-optimized human-assisted step in the process, we minimized the amount of coding and prior knowledge about the extracted property necessary to achieve a high recall and nearly perfect precision. A modest sized database of up to around 1000 entries can be extracted in around one workday with this method. The method is assessed vs. ground truth on a bulk modulus database and then applied to construct curated database of critical cooling rate of metallic glasses.

\section{Datasets}
\label{sec:datasets}
Below, the details about the datasets are provided. As a result of this paper a high quality database of critical cooling rates for bulk metallic glasses has been curated, as well as a benchmark-only dataset - the bulk modulus dataset, which was used to assess the model. Information on accessing the datasets can be found in Sec.~\ref{sec:data_availability}. It is important to note that we only used papers for which a full text is available in a text (xml) format. The cut-off date for availability of full texts of papers varies from journal to journal, but is usually around the mid 2000s. Fortunately, however, despite not having access to older papers, a significant amount of valuable or relevant older data is likely gathered as well, as that data is often repeated and referred to in more recent papers, which is then subsequently extracted with our method.

\subsection{Bulk modulus sentences}
\label{sec:dataset_bulkmod}
The bulk modulus is a benchmark dataset of sentences. From over 10000 paper results of a search query \emph{''bulk modulus''+''crystalline''}, a subset of 100 papers from the first 6000 full-text papers available through ScienceDirect API was randomly selected. In the written text of these 100 papers, there are 18408 sentences in total, out of which 237 sentences mention the value of bulk modulus.
This sentence dataset is used as a benchmark for the classification so a human ground truth is extracted. To avoid excessive time spent establishing the human labeled ground truth this database uses only a very small fraction of the total available papers. Thus the bulk modulus sentences database is not nearly complete, and the human-assisted extraction of a final database is not performed.

For the zero-shot case (only step 1 and not step 2) the approach effectively has no training data and can just be assessed on the test data described above. However, when step 2 included the fine-tuning requires additional data (effectively a training data set). For this fine-tuning process an additional 339 positive and 484 negative sentences have been extracted from papers not included in the $100$ papers in the test set. These additional sentences are use to investigate how fine-tuning improves the model and plot learning curves (see Fig.~\ref{fig:learning}).

\subsection{Critical cooling rates for bulk metallic glasses}
\label{sec:dataset_critcool}
This dataset consists of data gathered from 668 papers based on a result of a search query ''bulk metallic glass''+''critical cooling rate'' from Elsevier's ScienceDirect API. These papers consisted of 107386 sentences, out of which 347 were identified as positive (containing values of critical cooling rates), after applying the workflow described in Sec.~\ref{sec:method}, including the optional Step 2 in order to provide best quality data. From these 347 sentences, 443 critical cooling rate data points (consisting of the material name, critical cooling rate value and units) were extracted and are collected as a final database presented. Additionally, that data was manually postprocessed to include only unique datapoints (removing duplicate results, i.e. the same values reported in multiple papers), remove those which included ranges or limits or values, or where the material's composition was not explicitly given, and unify the formatting of the materials compositions, which resulted in 211 unique datapoints. The total human time required for gathering this dataset did not exceed 5 hours.

\begin{acknowledgments}
\label{Sec:acknowledgments}
The research in this work was primarily supported by the National Science Foundation Cyberinfrastructure for Sustained Scientific Innovation (CSSI) Award No. 1931298. Additional support for engaging undergraduates in research was provided by the National Science Foundation Training-based Workforce Development for Advanced Cyberinfrastructure (CyberTraining) Award No. 2017072. We thank Shishmitha Adusumilli, Harmon Bhasin, and Shanchao Liang for their participation in this project.
\end{acknowledgments}

\section{Contributions}
\label{sec:contributions}
M. P. P. conceived the study, performed the modeling, tests and prepared/analyzed the results, S. M., J. Z., J. W., S. W. and A. D. H. performed tests and assessed methods as a part of the Informatics Skunkworks program for undergraduate researchers, A. L. post-processed and extracted data, D. M. guided and supervised the research. Writing of the manuscript was done by M. P. P. and D. M.. All authors read, revised and approved the final manuscript.

\section{Methods}

As with any machine learning model, there are hyperparameters that may be optimized. Our experience showed that there is very little to be gained by performing the costly optimization, and throughout the paper we used default recommended values for all models. In OpenAI \texttt{GPT-3 davinci} we used the default values for fine-tuning, and when using both the pretrained \texttt{text-davinci-002/003} and fine-tuned \texttt{davinci} we set the frequency and presence penalties as well as temperature to $0$. The fine-tuning of \texttt{bart} and \texttt{DeBERTaV3} was performed with default recommended values too, which is a learning rate of $2e-5$, batch size of $16$, $5$ epochs, and $0.01$ weight decay. Full and detailed input files can be found in \cite{figshare}. Python codes were executed with \texttt{Python} ver. $3.10.6$.
For zero-shot classification with OpenAI models, the model's response was limited to a single token to facilitate a yes/no answer and preserve resources by cutting off further completion. A \texttt{0613} (June 13th 2023) snapshots of OpenAI \texttt{chat} models, \texttt{GPT-3.5 (turbo)} and \texttt{GPT-4}, were used, with an empty system message. For each sentence classification a separate chat was initiated.

\subsection{Definition of statistical quantities}
\label{sec:stats}

\noindent\emph{True positive (TP)} - a sentence containing numerical data for a given property.\\
\emph{True negative (TN)} - a sentence not containing numerical data for a given property.\\
\emph{False positive (FP)} - a sentence not containing the numerical data for a given property but is identified as one that does.\\
\emph{False negative (FN)} - a sentence containing numerical data for a given property but is identified as one that does not.\\

Precision:
\begin{equation}
Precision=\frac{TP}{TP+FP}
\end{equation}
Recall (True Positive Rate):
\begin{equation}
Recall=\frac{TP}{TP+FN}
\end{equation}
False Positive Rate (FPR):
\begin{equation}
FPR=\frac{FP}{FP+TN}
\end{equation}
F1 score:
\begin{equation}
FPR=\frac{2TP}{2TP+FP+FN}
\end{equation}

\subsection{Data Availability}
\label{sec:data_availability}
The databases curated as a result of this paper, all datasets used in the assessment of the method, as well as the codes and software used in this paper are available on figshare \cite{figshare}: \hyperlink{https://doi.org/10.6084/m9.figshare.21861948}{https://doi.org/10.6084/m9.figshare.21861948}. The codes are included for full transparency, but were developed for internal use only, so they contain very limited error handling, and the authors do not guarantee that they will work universally on every system. All parameters used for the model fine-tuning and zero-shot classification can be found in the codes.

\subsection{Competing Interest}
The authors declare no competing interest.

\end{document}